\documentclass[twocolumn]{jpsj2} 

\title{
Three-orbital Kondo effect in single quantum dot system with plural electrons}

\author{
Tomoko \textsc{Kita}$^{1}$
\thanks{E-mail address: kita@tp.ap.eng.osaka-u.ac.jp}, 
Rui \textsc{Sakano}$^{1}$
\thanks{Present address: Department of Applied Physics, University of Tokyo, 
Hongo 7-3-1, Bunkyo-ku, Tokyo 113-8656.}, 
Takuma \textsc{Ohashi}$^{2}$, and 
Sei-ichiro \textsc{Suga}$^{1}$ 
}

\inst{
$^{1}$Department of Applied Physics, Osaka University, Suita, Osaka 565-0871 
\\
$^{2}$Department of Physics, Osaka University, Toyonaka, Osaka 560-0043
}

\abst{
We study the Kondo effect and related transport properties 
in orbitally degenerate vertical quantum dot systems 
with plural electrons. 
Applying the non-crossing approximation to the three-orbital Anderson impurity
model with the finite Coulomb interaction and Hund-coupling,
we investigate the magnetic-field dependence of
the conductance and thermopower. 
We also introduce an additional orbital splitting 
to take account of the realistic many-body effect in 
the vertical quantum dot system. 
It is clarified 
how the three-orbital Kondo effect influences the transport properties 
via the modulation of the Kondo temperature and 
unitary limit of transport quantities 
due to the change of the symmetry in the system. 
}

\kword{quantum dot, Kondo effect, orbital degrees of freedom, thermopower, non-crossing approximation}

\begin{document}
\maketitle

\section{Introduction}
Since it was found that the Kondo effect significantly influences
transport properties of quantum dot (QD) systems 
electron correlations in such nanoscale systems 
have attracted much attention. 
Although the Kondo physics has been investigated 
for more than forty years and the mechanism has been already understood 
\cite{kondo64,hewson93,JPSJ},
the effect in nanoscale systems has received a lot of renewed interest. 
Artificial nanoscale systems, like QD systems, have a lot of controllable 
parameters, which makes it possible to explore a new aspect of the Kondo physics. 
For instance, recent nanofabrication technology has enabled us to externally 
tune orbital properties of QD systems. 
In vertical QD 
\cite{tarucha96_97,tarucha00,tarucha00_PRL,kouwenhoven01,austing99,tokura01}
and the carbon nanotube QD 
\cite{moriyama05,cobden02,j_herrero04,j_herrero05_PRL,white98} 
systems, effective orbital degeneracy is generated by 
their highly symmetric shapes of the electron confinement potential. 
Also in multiple QD systems, it was found that 
pseudo orbital states are induced by 
the interplay between electrons in QDs 
\cite{wilhelm02,sun02,izumida00,dong02prb,nishikawa06,mravlje06,borda03,sakano05,galpin05,lipinski06,boese02,lopez05}. 
In these systems, 
the characteristic transport properties due to the orbital Kondo effect 
have been studied both experimentally and theoretically.

Here, we focus on the vertical QD, 
where electrons are confined in a two dimensional (2D) harmonic potential. 
It yields multiply degenerate single-particle levels called the
Fock-Darwin states \cite{fock28,darwin30}, 
which are regarded as effective orbital degrees of freedom. 
Recent progress in experimental techniques of 
tuning the 2D confinement potential allows us to 
construct even triply degenerate orbital states 
and control their degeneracy by applying magnetic fields. 
This system provides a well-controlled quantum device 
to study the transport properties due to 
the three-orbital Kondo effect. 

The orbital Kondo effect has been formerly investigated in the heavy
fermion system and a number of the studies have been done 
\cite{kuramoto83,zhang83,coleman83,maekawa85}. 
In QD systems, the two-orbital Kondo effect 
such as the {\it SU}(4) Kondo effect
\cite{borda03,sakano05,galpin05,lipinski06,boese02,lopez05,j_herrero05,makarovski07PRL,makarovski07PRB,choi05,lim06,anders07,sasaki04,zhuravlev04,eto05,sakano06,inoshita93} 
and the singlet-triplet Kondo effect
\cite{sasaki00,eto00_01,pustilnik00_01,izumida01,hofstetter02,pustilnik03,imai01,eto02,pustilnik00,izumida98,schmid00,maurer99,nygard00}
have been intensively studied.
However, there are a few studies on the orbital Kondo effect 
in three- or more-orbital models 
and the transport properties are not apparent.
In addition, 
it is desirable to investigate 
not only the conductance but also the thermopower 
\cite{scheibner05,boese01,kim02_03,krawiec06,donabidowicz07,krawiec07,sakano07,chen08,koshibae01}, 
which gives complementary information on the QD electronic state 
to the conductance. 

In this paper, we study the three-orbital Kondo effect and 
transport properties in a single vertical QD system. 
By applying the non-crossing approximation (NCA) 
\cite{keiter84,pruschke89,bickers87,lombardo05} 
to the three-orbital Anderson impurity model 
with the finite Coulomb interaction and Hund-coupling, 
we calculate the local density of states (DOS) of the QD, 
conductance and thermopower. 
In particular, we investigate the modulation of the Kondo effect by tuning external magnetic fields, and how it influences transport properties observed experimentally. 
To consider an experimental environment in the vertical QD, 
we introduce the $\lambda$-perturbation \cite{matagne02_realistic}, namely, 
a deviation from the ideal Fock-Darwin state. 
We find that three orbital degrees of freedom and 
the effects of the $\lambda$-perturbation induce some noticeable properties 
in the conductance and thermopower, 
which do not appear in the ordinary spin and two-orbital cases 
studied previously. 

This paper is organized as follows.
In \S\ref{sec:model}, we introduce the model and
briefly mention the NCA approach. 
In \S\ref{sec:result}, we present our numerical results, and 
discuss the magnetic-field dependence of the conductance and thermopower 
in the presence of the $\lambda$-perturbation. 
A brief summary is given in \S\ref{sec:sum}.

\section{Model and Method}
\label{sec:model}
In this section, we introduce our vertical QD system 
with orbital degrees of freedom, 
which is modeled as the three-orbital Anderson impurity Hamiltonian. 
Then, we outline the theoretical framework to treat the Kondo effect 
and to derive the linear conductance and thermopower. 

\subsection{Three-orbital quantum dot}
\begin{figure}[tbp]
\begin{center}
\includegraphics[width=.25\textwidth]{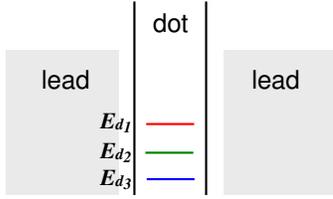}
\caption{
(Color online) 
Schematic diagram of the energy states in our QD system. 
The QD with three orbitals is coupled to two leads. 
}
\label{fig:3orb_singleQD}
\end{center}
\end{figure}

Let us consider a single QD with plural electrons. 
The single-particle energy levels of the QD can be described 
as the Fock-Darwin states. 
As seven, eight, or nine electrons occupy 
the Fock-Darwin states at zero field 
successively from the lowest energy level, 
six electrons form the closed-shell and 
the extra electrons occupy the triply degenerate single-particle level, 
which can be regarded as three orbital degrees of freedom. 
The electrons in the triply degenerate orbitals 
contribute to the three-orbital Kondo effect and transport properties. 
Therefore, we use the model illustrated in Fig. \ref{fig:3orb_singleQD}, 
which represents the QD with the three orbitals coupled to two leads. 
The system with four (five) electrons in three orbitals 
corresponds to that with two (one) electrons via electron-hole symmetry. 

In the vertical QD system, 
a magnetic field lifts the orbital degeneracy. 
The single-particle energy levels of three orbitals are 
split into three distinguished levels, $E_{d_1}$, $E_{d_2}$ and $E_{d_3}$, 
by the orbital Zeeman effect. 
Note that the Zeeman splitting of spin states is much smaller than the orbital one, 
so that we can ignore the spin splitting. 
The orbital splitting $\Delta_B$ is usually proportional to 
strength of a magnetic field $B$. 
Therefore, it allows us to study the magnetic-field dependence 
by analyzing the $\Delta_B$ dependence of our system. 
Then, the three orbital levels of the QD under a magnetic field can be written as 
\begin{eqnarray}
E_{d_l}&=& \varepsilon _c + (2-l)\Delta _{B} \nonumber \\
       && + \Delta _{\lambda}(2 \delta _{l,2}- \delta _{l,1}- \delta _{l,3})/3 \label{eq:energylevel} \quad (l=1, 2, 3), 
\end{eqnarray}
where $\varepsilon_c$ denotes the center of the energy levels. 

\begin{figure}[tbp]
\begin{center}
\includegraphics[width=.45\textwidth]{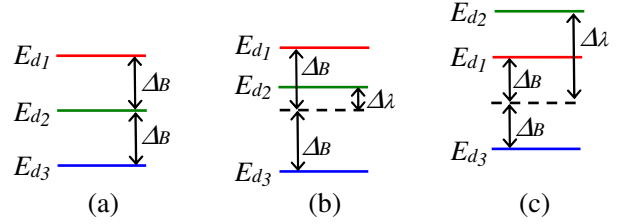}
\caption{
(Color online) 
Level-splitting scheme of three orbital-levels in the QD.
(a) $\Delta_{\lambda}=0$,
(b) $\Delta_{\lambda} \ll \Delta_B$,
and (c) $\Delta_{\lambda} \gg \Delta_B$.
}
\label{fig:1}
\end{center}
\end{figure}

To consider the experimental environment in vertical QD systems, 
we here introduce an additional orbital splitting $\Delta_{\lambda}$, 
corresponding to deviations from an ideal Fock-Darwin state. 
According to ref. \citen{matagne02_realistic}, 
this perturbation called the $\lambda$-perturbation 
gives a correction term to the parabolic confinement potential 
and lifts the degeneracy between two degenerate orbitals 
and the other orbital. 
In addition to the deviation from the parabolic potential, 
deformation of the QD and a many-body effect among electrons 
induce the same perturbation effect, 
resulting in the orbital splitting $\Delta_{\lambda}$ \cite{tarucha91}.
In fact, this $\lambda$-perturbation is observed experimentally. 
We show schematic diagrams of these orbital configurations in Fig. \ref{fig:1}. 

To consider the Kondo effect, 
we describe our QD system by the three-orbital Anderson impurity model, 
\begin{eqnarray}
{\cal H} &=& {\cal H}_c + {\cal H}_d+{\cal H}_t, 
\label{eq:model} \\
{\cal H}_c &=& \sum _{ k l \sigma } 
            \varepsilon_k c_{k l \sigma }^{\dagger } c_{k l \sigma}, 
\label{eq:Hc} \\
{\cal H}_d &=& \sum _{l \sigma}
            E_{d_l} d_{l \sigma}^{\dagger }d_{l \sigma} 
\nonumber\\
&& + U \sum _{l \sigma \neq l^\prime \sigma^{\prime}}
              n_{d_{l \sigma}}n_{d_{l^\prime \sigma^{\prime}  }}
       -  J \sum _{l\neq l^\prime} 
              \textbf{S}_{d_l} \cdot \textbf{S}_{d_{l^\prime}},
\label{eq:Hd} \\
{\cal H}_t &=& \sum _{ k l \sigma} 
            V_{kl\sigma} \left(c_{k l \sigma }^{\dagger }d_{l \sigma}+ \mbox{H.c.} \right). 
\label{eq:Ht}
\end{eqnarray}
Here, ${\cal H}_c$ represents the electronic states in the leads and
$c_{k l \sigma }^{(\dagger)}$ annihilates (creates) a conduction electron
with the dispersion $\varepsilon_k$ and spin $\sigma$ 
in the orbital $l$. 
The QD part is described as ${\cal H}_d$, 
where $d_{l \sigma}^{(\dagger)}$ annihilates (creates) a localized electron 
in the QD with spin $\sigma$ in the orbital $l$,
$n_{d_{l\sigma}}=d_{l \sigma}^\dag d_{l \sigma}$, and 
$\textbf{S}_{d_l}=\sum _{\sigma \sigma^\prime} d_{l \sigma}^\dagger (\boldsymbol{\sigma}_{\sigma \sigma^\prime}/2) d_{l \sigma^\prime}$ 
is the spin operator for a localized electron in the orbital $l$ with 
$\boldsymbol{\sigma}$ being the Pauli matrices.
The Coulomb interaction and the ferromagnetic Hund-coupling 
are expressed by $U$ and $J$, respectively. 
The tunnel process between the leads and the QD is written as 
${\cal H}_t$. 
We assume that the orbital states in the QD
hybridize with the corresponding conduction channels
in the leads for simplicity.
Although the assumption of multiple conduction channels is nontrivial, 
it is known that this is relevant for certain
systems, {\it e.g.} vertical QD systems and carbon nanotube QD systems
\cite{sasaki04,eto05,j_herrero05}. 

\subsection{Non-crossing approximation}
In order to study the three-orbital Kondo effect 
for the model eq. (\ref{eq:model}), 
we make use of the NCA. 
This method is known to give physically sensible results at temperatures around and higher than the Kondo temperature
\cite{keiter84,pruschke89,bickers87,lombardo05}. 
In fact, it has been successfully applied to the Ce and Yb impurity problem, 
for which orbital degrees of freedom play an important role \cite{kuramoto83,zhang83,coleman83,maekawa85}. 
In particular, 
we use the NCA to treat the finite Coulomb interaction and Hund-coupling.
We outline the procedure of the NCA below. 

We first diagonalize the QD part of the Hamiltonian ${\cal H}_d$.
It can be expressed in terms of the Hubbard operator $X_{mn}=|m\rangle \langle n|$ as
\begin{eqnarray}
{\cal H}_d=\sum _{m=1}^{64}E_m X_{mm},
\end{eqnarray}
where $E_m$ and $|m\rangle $ denote the $m$-th eigenenergy and 
eigenstate of ${\cal H}_d$, respectively.
The mixing term ${\cal H}_t$ is also rewritten in terms
of the Hubbard operators as 
\begin{eqnarray}
{\cal H}_t = \sum_{k l \sigma} \sum_{mn} V_{kl\sigma} M_{mn}^{l \sigma}
(c_{k l \sigma}^{\dagger}X_{nm}+ \mbox{H.c.}),
\end{eqnarray}
where the matrix element $M_{mn}^{l \sigma}$ is given by
$M_{mn}^{l \sigma}=\langle m|d_{l \sigma}^{\dagger}|n\rangle$. 
For each eigenstate $|m\rangle$, the ionic propagator 
and the corresponding spectral function can be defined as 
\begin{align}
R_{m}(\omega )&=\frac{1}{\omega -E_{m}-\Sigma_{m}(\omega )},
\label{eq:propagator} \\
\rho_{m}(\omega )&=-\frac{1}{\pi }{\rm Im}R_{m}(\omega +{\rm i}\eta). 
\end{align}
We evaluate the self-energy $\Sigma_{m}(\omega )$ by 
a self-consistent perturbation theory up to the
second order of $H_{t}$. 
The self-energy is thus obtained as 
\begin{align}
\Sigma_{m}(\omega )=\sum_{n,l,\sigma} \bigg[ &
\left|M_{nm}^{l \sigma}\right|^{2}
\int d \varepsilon 
A_{c_{l \sigma}}(\varepsilon ) R_{n}(\omega +\varepsilon ) f(\varepsilon ) 
\nonumber \\
+ & \left|M_{mn}^{l \sigma}\right|^{2}\int d \varepsilon 
A_{c_{l\sigma}} (\varepsilon ) R_{n}(\omega -\varepsilon ) f(-\varepsilon ) \bigg],
\label{eq:selfenergy}
\end{align}
where 
$f(\varepsilon)$ 
is the Fermi distribution function and 
$A_{c_{l\sigma}}(\varepsilon )=\sum_{k} \left|V_{k l\sigma} \right|^2 \delta(\varepsilon-\varepsilon_k)$. 
Equations 
(\ref{eq:propagator}) and (\ref{eq:selfenergy}) 
complete the self-consistent loop. 
The local DOS in the QD is obtained as 
\begin{align}
\rho_{l \sigma}(\omega ) &
= \frac{1}{Z_{\textrm{loc}}}
\sum_{mn} \left|M_{mn}^{l \sigma }\right|^{2}
\int d\varepsilon 
e^{- \varepsilon / T} \nonumber \\
& \times
\left[ \rho_m (\varepsilon )\rho_n (\varepsilon-\omega)
+\rho_m (\varepsilon+\omega )\rho_n (\varepsilon ) \right]. 
\label{eq:rho_d}
\end{align}
Here, $Z_{\textrm{loc}}$ is the partition function of the QD, 
\begin{eqnarray}
Z_{\textrm{loc}}=\sum_{m}\int d\omega 
e^{- \omega / T}\rho_{m}(\omega). 
\end{eqnarray}
The total number of electrons in the QD is also calculated as 
\begin{eqnarray}
n_{tot} = \sum_{l \sigma} \int d\omega \rho _{l \sigma}(\omega ) f (\omega). 
\end{eqnarray}

The non-equilibrium Green's function technique allows us to study various transport properties, which gives the expressions for the bias-linear conductance \cite{meir92_94} and the temperature-linear thermopower \cite{dong02} as
\begin{eqnarray}
G &=& - \frac{\pi e^2}{h}
\sum _{l \sigma} \int d \omega f'(\omega)
 \Gamma _{l \sigma}(\omega) \rho _{l \sigma}(\omega),
\label{G}
\\
S &=& -\frac{1}{eT} 
     \frac{ \sum _{l \sigma} \int d \omega f'(\omega)
      \omega \Gamma _{l \sigma}(\omega) \rho _{l \sigma}(\omega)
          }
          {\sum _{l \sigma} \int d \omega f'(\omega)
        \Gamma _{l \sigma}(\omega) \rho _{l \sigma}(\omega)
          }
\label{S}
\end{eqnarray}
with the energy derivative of the Fermi distribution function, $f'(\omega)=df(\omega)/d\omega$.
Here, $\Gamma_{l \sigma}(\omega)$ denotes the strength of the hybridization
between the conduction electron and the localized electron in the QD,
described by 
$\Gamma_{l \sigma}(\varepsilon)=\pi A_{c_{l\sigma}}(\varepsilon )$.
In the following calculations,
for simplicity, 
we adopt the wide band limit
and assume the spin and orbital independence, 
which allows us to rewrite it as a constant $\Gamma$.
We will take $\Gamma$ in units of energy in the following sections. 
By numerically evaluating the DOS of the QD, 
we investigate the conductance and the thermopower of the QD system.

Before turning to discussions on the numerical results, 
we comment on the applicability of the NCA to systems at lower temperatures. 
As mentioned above, the NCA method is valid around or higher than the Kondo temperature, 
but it gradually breaks down for the Fermi liquid regime at lower temperatures. 
Since the Kondo temperature changes with the orbital configuration, 
our calculation of the level-splitting dependence 
for fixed temperatures may be less reliable. 
However, the qualitative behavior of transport properties 
with varying the orbital level-splitting 
is described well by the NCA even at lower temperatures. 
We will return to this point later. In the end of the next section, 
we will complementally investigate the temperature dependence of transport properties. 

\section{Numerical Results}
\label{sec:result}
In this section, we discuss how magnetic fields influence the three-orbital 
Kondo effect and transport properties in the QD systems. 
We investigate the situations where the QD contains 
one electron, two electrons, and three electrons, in each subsection. 
As mentioned in the previous section, 
they correspond to the orbital Kondo effects observed experimentally 
in the vertical QD with seven, eight, and nine electrons, respectively. 
We calculate the conductance and thermopower as a function of the 
orbital splitting $\Delta_B$, which correspond to the magnetic-field 
dependence in a real QD under $\lambda$-perturbation $\Delta_\lambda$. 
We show that the behavior of these transport properties are well 
understood by the three orbital Kondo effects with orbital splitting. 
The modulation of the Kondo temperature by several types of the orbital 
splitting is discussed in the following subsections. 
The NCA calculations are done at a mainly fixed finite temperature 
but at as low temperature as possible, 
where NCA gives reliable results for whole parameter region of 
$\Delta_B$ and $\Delta_\lambda$, as discussed below.

\subsection{Three-orbital Kondo effect with one electron; \\ $n_{tot}=1$}
\begin{figure}[tbp]
\begin{center}
\includegraphics[width=.9\linewidth]{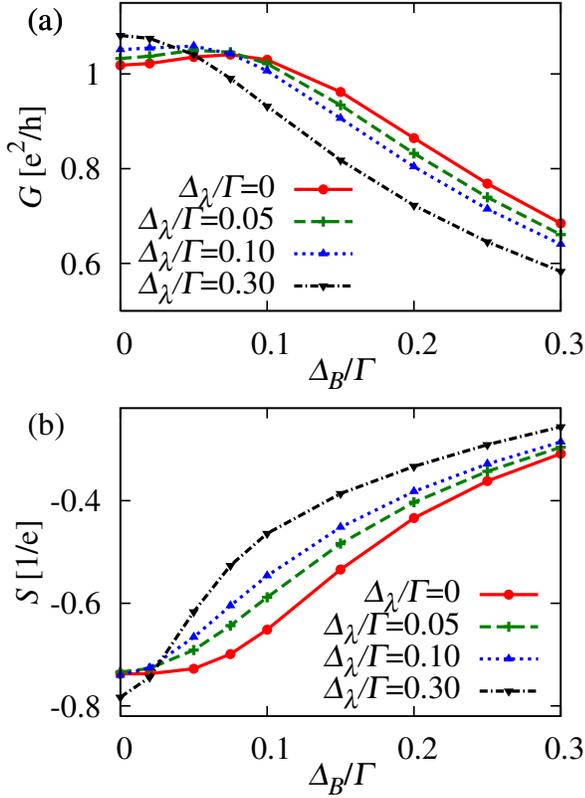}
\caption{
(Color online)
(a) The conductance $G$ and (b) the thermopower $S$
as a function of magnetic fields $\Delta_{B}$ 
for $\Delta_{\lambda}/\Gamma=0, 0.05, 0.10, 0.30$
(solid, dashed, dotted, dash-dotted line).
The parameters are set as $U/\Gamma=10$, $J=0$, and $T/\Gamma=0.016$.
}
\label{fig:1eGS}
\end{center}
\end{figure}

Let us start with
the three-orbital Kondo effect with one electron ($n_{tot}=1$):
The center of the energy levels in eq. (\ref{eq:energylevel}) is fixed at $\varepsilon_c = -U/2$.
In this subsection, 
we set parameters as the Coulomb interaction $U/\Gamma=10$ 
and the temperature $T/\Gamma=0.016$. 
The magnetic-field $\Delta_{B}$ dependence of 
the conductance $G$ and the thermopower $S$ are shown in Fig. \ref{fig:1eGS} for several choices of the $\lambda$-perturbation $\Delta_\lambda$.
Since we here consider the regime
that one electron is contained in the three-fold orbital of the QD,
the Hund-coupling which works between electrons 
hardly affects the Kondo effect and transport properties,
so that we set $J=0$ \cite{zhuravlev04,sakano07}.

In the absence of the $\lambda$-perturbation, $\Delta_\lambda=0$,
we find that the conductance $G$ plateaus at small $\Delta_B$.
This characteristic behavior in the conductance indicates the existence of the three- or more-orbital Kondo effect, which can be explained by the unitary conductance and the effective Kondo temperature of the dominant Kondo state.
For small magnetic fields, the {\it SU}(6) Kondo effect is dominant due to the three-fold orbital state, while for large magnetic fields, the orbital degeneracy is smeared and the {\it SU}(2) Kondo effect due to spin is dominant.
Therefore, at absolute zero,
the conductance monotonously increases from $G=3e^2/2h$ with fields,
and approaches $G=2e^2/h$ \cite{sakano06}.
However, the applied magnetic field simultaneously decreases the effective Kondo temperature, which results in the suppression of the Kondo effect at finite temperatures.
Here, the order of the Kondo temperature is estimated to be $T_K^{SU(6)} \sim 0.1\Gamma $ and $T_K^{SU(2)} \sim 0.01 \Gamma$ from the resonance width of the computed DOS.
The competition between the enhancement and the suppression of the conductance due to the Kondo effect at finite temperatures gives rise to the conductance plateau around $\Delta_B \sim T_K^{SU(6)}$.
For large $\Delta_{\lambda}$,
the conductance for not only both the zero magnetic-field limit $\Delta_B=0$ ({\it SU}(4) Kondo is dominant) and the large magnetic-field limit $\Delta_B= \infty$ ({\it SU}(2) Kondo is dominant), but also the intermediate region $\Delta_B = finite$ are given by $G=2e^2/h$ at absolute zero. 
On the other hand, 
the decrease of the effective Kondo temperature is given by $T_K^{eff} \sim 1/\Delta_B$ for large $\Delta_B$.
Thus, it results in the monotonous decrease of the conductance with $\Delta_B$.
The conductance behavior obtained in this study is qualitatively 
consistent with that 
for the Anderson impurity model in the strong coupling limit 
($U \to \infty $) \cite{sakano06}.
Since the {\it SU}(4) Kondo temperature $T_K^{SU(4)}(\sim 0.05 \Gamma)$ is smaller than the {\it SU}(6) Kondo temperature $T_K^{SU(6)}$, the conductance decreases faster with magnetic fields $\Delta_B$ for larger $\Delta_{\lambda}$.
Around $\Delta_{\lambda} = T_K^{SU(6)} \sim 0.1 \Gamma$,
the magnetic-field dependency of the conductance behavior crossovers
from the three-orbital Kondo regime to the two-orbital one.

Similarly to the conductance behavior,
we can observe the dominant Kondo state in the thermopower behavior.
The computed thermopower $S$ as a function of applied magnetic fields $\Delta_B$ for several choices of $\Delta_{\lambda}$ are shown in Fig. \ref{fig:1eGS}(b).
One can see that the thermopower takes negative values and approaches zero with increasing $\Delta_B$.
This means two things. First one is that the Kondo resonance peak is located above the Fermi level, which indicates the orbital Kondo effect.
Secondly, in the presence of magnetic fields, 
the Kondo resonance approaches the Fermi level and is suppressed, 
which indicates that the {\it SU}(2) Kondo effect occurs.
Then, the thermopower for large $\Delta_{\lambda}$ approaches zero faster than for small $\Delta_{\lambda}$ as increasing $\Delta_B$, because of $T_K^{SU(4)} < T_K^{SU(6)}$.
Since the thermopower due to the Kondo resonance for more orbital degeneracy is smaller,
the thermopower for small $\Delta_{\lambda}$ takes smaller values 
at zero field $\Delta_B=0$.
Around $\Delta_{\lambda} = T_K^{SU(6)} \sim 0.1 \Gamma$,
the magnetic-field dependency of the thermopower also crossovers from the three-orbital Kondo regime to the two-orbital one.

Note that this regime connects to $n_{tot}=5$ regime via the electron-hole transformation.
Therefore, in the $n_{tot}=5$ regime, the similar magnetic-field dependency of the transport with the opposite sign of the thermopower is expected.

\subsection{With two electrons; $n_{tot}=2$}
We now move to the two-electron regime $n_{tot}=2$:
The center of the energy levels in eq. (\ref{eq:energylevel}) is 
$\varepsilon_c = -3U/2$. 
Since plural electrons are contained in folding orbitals,
the Hund-coupling between two electrons plays an important role in the Kondo effect and the transport.
The similar Kondo effect in QD systems with a two-fold orbital state has been observed as the singlet-triplet Kondo effect
\cite{eto00_01,pustilnik00_01,izumida01,hofstetter02,pustilnik03,imai01,eto02,pustilnik00,izumida98}.
Therefore, we believe that it is interesting to discuss how the third orbital affects the Kondo effect and transport properties, comparing two- and three- orbital Kondo effect.
For the purpose,
we set parameters as $U/\Gamma=10$, $J/\Gamma=0.5$, and $T/\Gamma=0.025$ in this subsection.
We note that this regime also connects to $n_{tot}=4$ regime via the electron-hole transformation and the difference is only the sign of the thermopower.
\begin{figure}[tbp]
\begin{center}
\includegraphics[width=.9\linewidth]{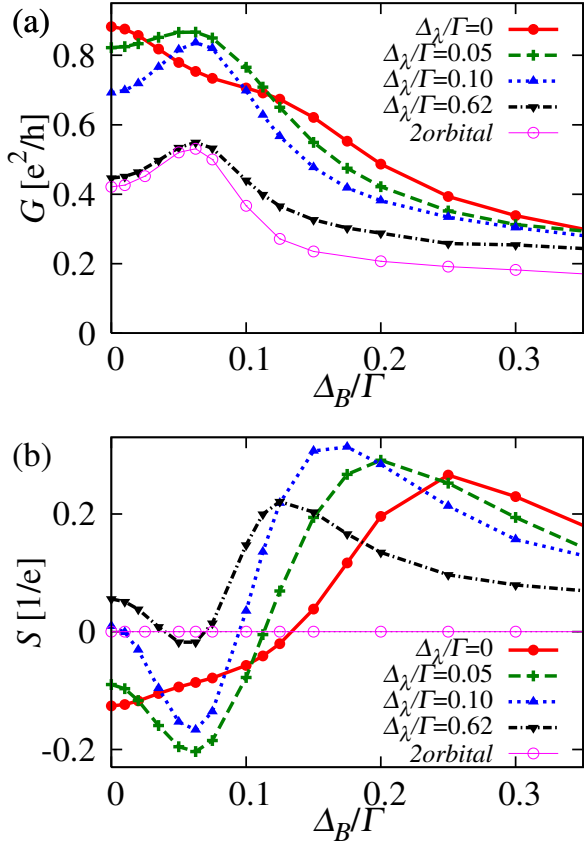} \\
\caption{(Color online) 
(a) The conductance $G$ and (b) the thermopower $S$
as a function of $\Delta_{B}$ for
$\Delta_{\lambda}/\Gamma=0, 0.05, 0.10, 0.62$
(solid, dashed, dotted, dashed-dotted line).
We also show the results for the two-orbital case,
$E_{d_1}-E_{d_2}=2\Delta_{B}$ (thin solid line),
to compare with results in the large limit of $\Delta_{\lambda}$.
The parameters are set as $U/\Gamma=10, J/\Gamma=0.5$, and $T/\Gamma=0.025$.
}
\label{fig:3}
\end{center}
\end{figure}

In Fig. \ref{fig:3}, 
we show the computed conductance $G$ and thermopower $S$ as a function of magnetic fields $\Delta_{B}$ with several choices of $\Delta_{\lambda}$. 
As described in Fig. \ref{fig:1}(b), 
the orbital level $E_{d_2}$ 
is shifted by $\Delta_{\lambda}$ even at zero field. 
In large $\Delta_{\lambda}$ limit, 
it is expected that the system approaches the two-orbital system, 
where $E_{d_1}-E_{d_3}=2\Delta_{B}$ 
as described in Fig. \ref{fig:1}(c). 
For comparison, 
we also show the results for the two-orbital case 
$E_{d_1}-E_{d_2}=2\Delta_{B}$.

\subsubsection{Zero $\lambda$-perturbation; $\Delta_{\lambda}=0$}
We first address the case that the $\lambda$-perturbation is negligibly small, 
$\Delta_{\lambda}=0$,
where only the magnetic field splits the energy levels by $\Delta_{B}$, 
as shown in Fig. \ref{fig:1}(a).
\begin{figure}[tbp]
\begin{center}
\includegraphics[width=\linewidth]{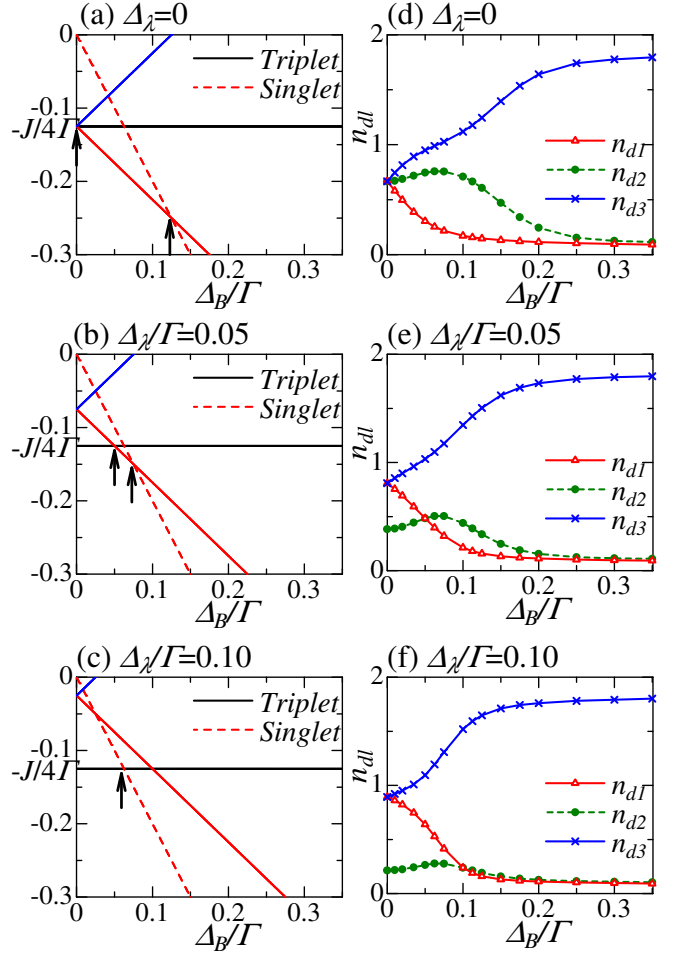} \\
\caption{(Color online) 
(a)-(c): The eigenenergies of $H_{d}$ 
as a function of $\Delta_{B}$ are described 
for $\Delta_{\lambda}/\Gamma=0$, $0.05$ (small $\Delta_{\lambda}$),
$0.10$ (large $\Delta_{\lambda}$), respectively.
To demonstrate the change of the lowest energy states,
the eigenenergies of three triplets (solid lines) and
one singlet (dashed line) are shown.
Arrows denote
the points where the degeneracy of lowest energy state increases.
(d)-(f): The electron number $n_{d_l}$ in the orbital $E_{d_l}$
as a function of $\Delta_{B}$ are shown
for $\Delta_{\lambda}/\Gamma=0$, $0.05$ (small $\Delta_{\lambda}$),
$0.10$ (large $\Delta_{\lambda}$), respectively.
Other parameters are the same as in Fig. \ref{fig:3}.
}
\label{fig:n_d}
\end{center}
\end{figure}
\begin{figure}[tbp]
\begin{center}
\includegraphics[width=.7\linewidth]{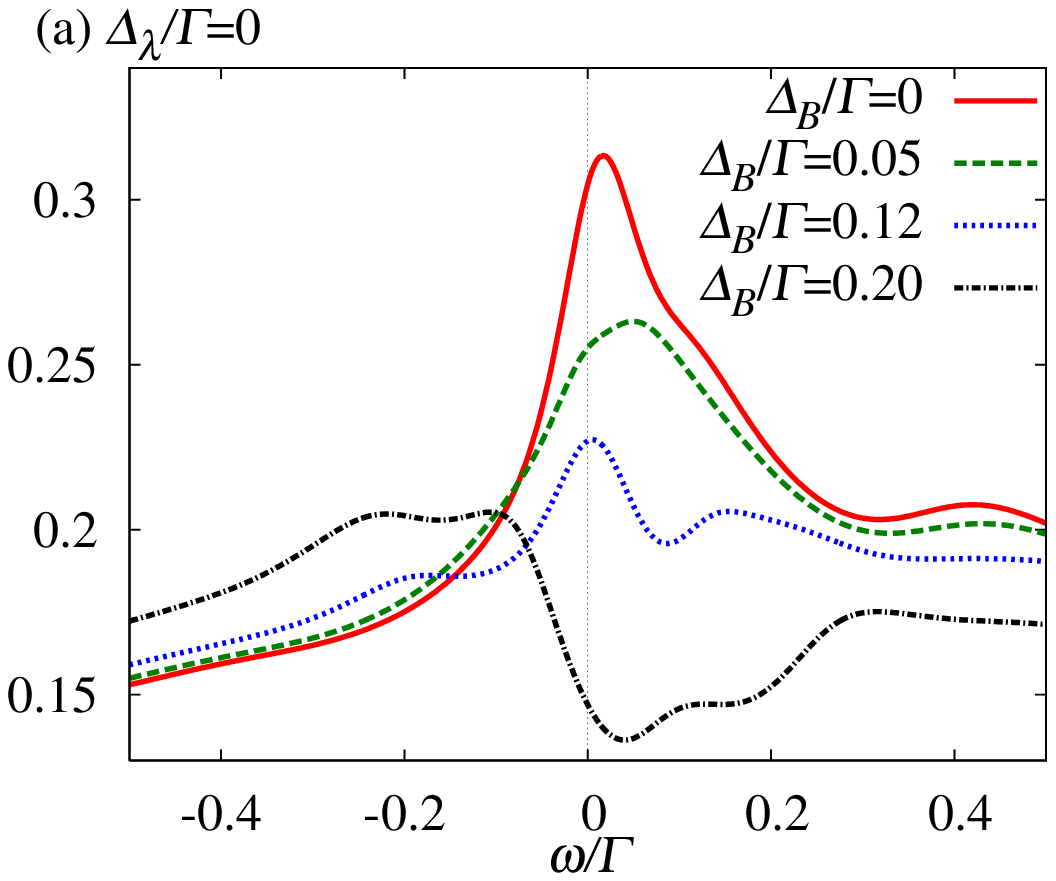}
\includegraphics[width=.7\linewidth]{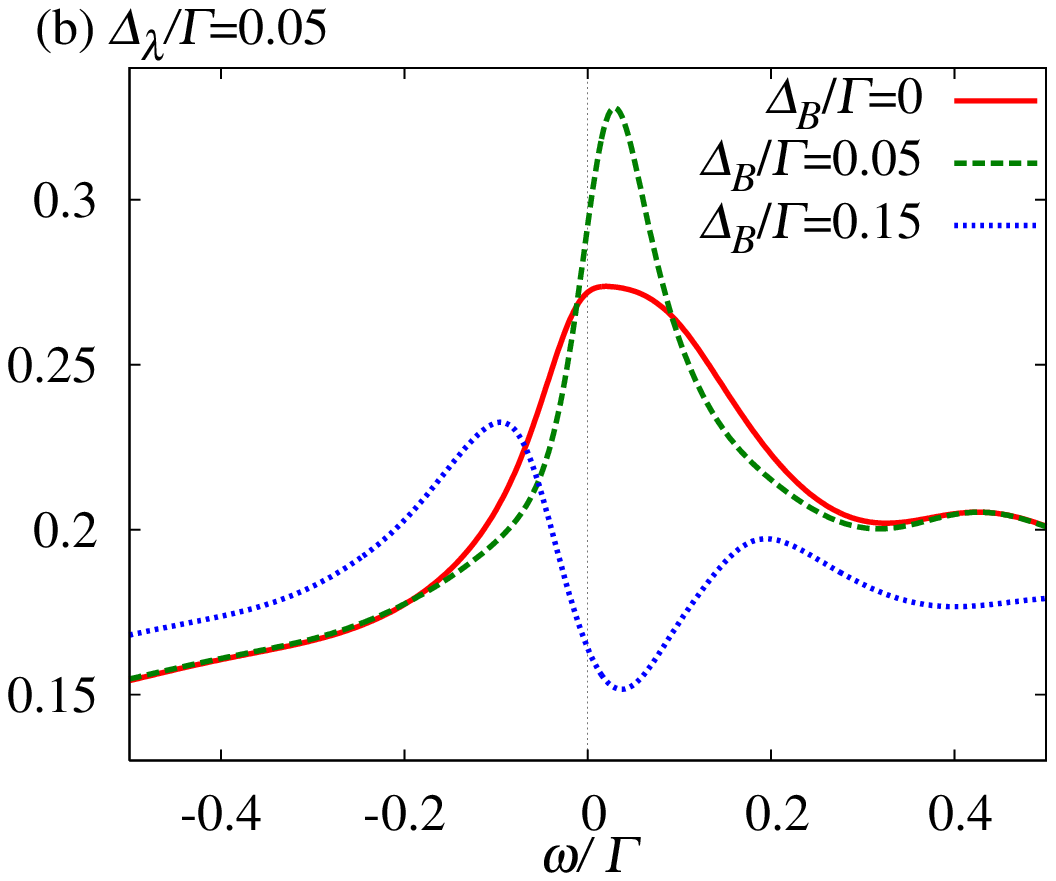}
\caption{(Color online) 
The local DOS $\rho(\omega)=\sum_{l \sigma}\rho_{l \sigma}(\omega)$
for (a) $\Delta_{\lambda}=0$, 
(b) $\Delta_{\lambda}/\Gamma=0.05$. 
The Fermi energy corresponds to $\omega=0$. 
Other parameters are the same as in Fig. \ref{fig:3}.
}
\label{fig:DOS}
\end{center}
\end{figure}

To understand how the Hund-coupling affects the Kondo effect and transport properties, it may be worthwhile to mention the well-known singlet-triplet Kondo effect or the two-orbital case
\cite{eto00_01,pustilnik00_01,izumida01,hofstetter02,pustilnik03,imai01,eto02,pustilnik00,izumida98}.
As shown in Fig. \ref{fig:3}(a),
the conductance $G$ for the two-orbital case takes a maximum
at $2\Delta_{B}/\Gamma \sim J/4\Gamma=0.125$.
At the point, the spin triplet and singlet states degenerate. 
This gives rise to the enhancement of the Kondo temperature 
and results in the increase of the conductance.

For the three-orbital case, it is also useful to consider the eigenenergies of the QD Hamiltonian $H_{d}$.
To demonstrate the change of the lowest energy states,
the eigenenergies of three triplets and one singlet
are shown as a function of magnetic fields $\Delta_{B}$
in Fig. \ref{fig:n_d}(a).
The arrows in Fig. \ref{fig:n_d} denote the points where the degeneracy of the QD ground state increases.
At $\Delta_{B}=0$, the lowest eigenstates have ninefold degeneracy.
The high degeneracy strongly enhances the Kondo effect and 
increases the conductance. 
By introducing the magnetic field $\Delta_{B}$, 
the ninefold degeneracy is lifted into three triplets. 
Therefore, in the small $\Delta_{B}$ regime, 
the conductance $G$ decreases as the magnetic field $\Delta_{B}$ increases. 
At $\Delta_{B}/\Gamma \sim J/4\Gamma=0.125$, 
the energies of the lowest triplet and the singlet states 
are degenerate, where it is expected that the conductance is enhanced 
by the singlet-triplet mechanism. 
However, the higher degeneracy at $\Delta_{B}=0$ 
obscures the enhancement of the conductance. 
Therefore, for the three-orbital case, 
the enhancement of the conductance caused by the singlet-triplet mechanism 
is not clear in contrast to the two-orbital case. 
Here, the order of the Kondo temperature for the 
triply degenerate orbital state 
is estimated to be 
$T_K^{\Delta_B=\Delta_\lambda=0} \sim 0.2\Gamma$ 
from the resonance width of the computed DOS shown in Fig. \ref{fig:DOS}(a).
Note that the enhancement of the conductance at $\Delta_{B} \sim J/4$ 
becomes a slight bit clearer 
in lower temperature, though it could not reach to the 
extent of the two-orbital case. 
The reason is that the conductance at $\Delta_{B}=0$, 
which reaches $G=9e^2/2h$ at zero temperature \cite{sakano06}, 
is larger than that of the two-orbital case 
because of more orbital degrees of freedom.

We can see a drastic difference between 
the two- and three-orbital cases 
also in the thermopower $S$, as shown in Fig. \ref{fig:3}(b). 
For the two-orbital case, 
the thermopower is zero irrespective of $\Delta_{B}$ 
because of the particle-hole symmetry. 
On the contrary, 
for the three-orbital case, the thermopower 
depends on $\Delta_{B}$. 
These results reflect the spectral profile of the DOS in the QD 
especially around the Fermi level. 
To put it more concretely, 
the peak position of the Kondo resonance 
mainly affects the sign of the thermopower. 
We show the DOS 
for the three-orbital case in Fig. \ref{fig:DOS}(a). 
At $\Delta_{B}=0$, 
two electrons occupy three degenerate orbitals, 
so that 1/3-filling, 
less than half-filling, is realized. 
Consequently, the Kondo resonance is located above the Fermi level, 
which leads to negative values of the thermopower. 
At $\Delta_{B}/\Gamma \sim J/4\Gamma=0.125$, 
where the singlet-triplet Kondo effect occurs, we find $S \sim 0$. 
We can see in Fig. \ref{fig:DOS}(a) that 
the singlet-triplet Kondo resonance 
at $\Delta_{B}/\Gamma = 0.125$ is located on the Fermi level. 
For further explanation, 
we show the electron number in each orbital $n_{d_l}$ 
as a function of $\Delta_{B}$ 
in Fig. \ref{fig:n_d}(d). 
At $\Delta_{B}/\Gamma \sim J/4\Gamma=0.125$, 
the electron number of the highest energy orbital $n_{d_1}$ 
is much fewer than others, $n_{d_2}, n_{d_3} \gg n_{d_1} \sim 0$. 
Therefore, the system behaves like the two-orbital system, 
the filling is effectively half, 
which results in the Kondo resonance located on the Fermi level. 
In general, we see that the thermopower reflects the effective filling 
through the peak position of the Kondo resonance. 

At $\Delta_{B}/\Gamma = 0.2$, 
the local DOS has a dip around the Fermi level. 
It indicates that 
the Kondo effect disappears for $\Delta_{B}/\Gamma \ge 0.2$. 
The behavior of the thermopower for $\Delta_{B}/\Gamma \ge 0.2$ 
does not reflect the Kondo resonance 
but only reflects the asymmetry originated from the local DOS.

\subsubsection{Effect of $\lambda$-perturbation; $\Delta_{\lambda}\neq 0$}
We next investigate the effect of the $\lambda$-perturbation 
$\Delta_{\lambda}$ on the Kondo effect and 
transport properties. 

As shown in Fig. \ref{fig:3}(a), 
$\Delta_{\lambda}$ obviously influences the magnetic-field dependence 
of the conductance $G$. 
For small $\Delta_{\lambda}$ ($\Delta_{\lambda}/\Gamma=0.05$), 
the conductance $G$ takes a broad maximum 
at $\Delta_{B}/\Gamma=0.05 \sim 0.06$. 
On the other hand, 
for large $\Delta_{\lambda}$ ($\Delta_{\lambda}/\Gamma=0.10, 0.62$), 
the conductance maximum becomes narrower 
and approaches the result of the two-orbital system 
with increasing $\Delta_{\lambda}$.
These results can be explained by eigenenergies of $H_{d}$ 
described in Fig. \ref{fig:n_d}. 

In Figs. \ref{fig:n_d}(b) and \ref{fig:n_d}(c), we show the results 
for $\Delta_{\lambda}/\Gamma=0.05$ and $\Delta_{\lambda}/\Gamma=0.10$ 
to study the effect of small and large $\Delta_{\lambda}$, respectively. 
For small $\Delta_{\lambda}$, 
as shown in Fig. \ref{fig:n_d}(b), 
we find that 
the two arrowed points, 
where the degeneracy increases, 
get closer to each other with $\Delta_{\lambda}$. 
Thus, the two enhanced points merge, which results in the broad maximum 
in the conductance $G$ for $\Delta_{\lambda}/\Gamma=0.05$ 
at $\Delta_{B}/\Gamma=0.05 \sim 0.06$. 
At $\Delta_B=0$, 
whereas the lowest eigenstates have ninefold degeneracy 
for $\Delta_{\lambda}=0$, 
it is reduced to the triplet for finite $\Delta_{\lambda}$. 
This leads to the decrease of the conductance at $\Delta_B=0$ 
with $\Delta_{\lambda}$. 
For large $\Delta_{\lambda}$ ($\Delta_{\lambda} \ge J/8=0.0625\Gamma$),
as shown in Fig. \ref{fig:n_d}(c),
it is same as the two-orbital case. 
Therefore, 
the conductance 
for $\Delta_{\lambda}/\Gamma=0.10, 0.62$
takes a maximum at $\Delta_{B}/\Gamma \sim J/8\Gamma =0.0625$
and approaches the two-orbital behavior with increasing $\Delta_\lambda$.

Let us turn to discuss the behavior of the thermopower.
The computed thermopower $S$ as a function of applied magnetic fields 
$\Delta_B$ for several choices of $\Delta_{\lambda}$ 
are shown in Fig. \ref{fig:3}(b). 
For $\Delta_{\lambda}/\Gamma=0.05$,
the absolute value of the thermopower takes a maximum
at $\Delta_{B}/\Gamma=0.05 \sim 0.06$,
where the conductance takes the broad maximum.
It is contrast to the behavior for $\Delta_{\lambda}=0$,
where $S \sim 0$ at the singlet-triplet point.
We study the effect of $\Delta_{\lambda}$
by the number of electrons in each orbital $n_{d_l}$, 
as shown in Fig. \ref{fig:n_d}(e). 
It is found that all three orbitals contain the nonzero number of electrons 
at $\Delta_{B}/\Gamma=0.05 \sim 0.06$. 
This means that all three orbitals are concerned with the Kondo effect.
Therefore, the system behaves as the 1/3-filling, which leads to
negative values of the thermopower. 
Increasing $\Delta_{\lambda}$ further,
for $\Delta_{\lambda}/\Gamma=0.10$ in Fig. \ref{fig:n_d}(f),
one can see that $n_{d_2}$ decreases with $\Delta_{\lambda}$. 
This means that only two orbitals, $E_{d_1}$ and $E_{d_3}$,
are concerned with the Kondo effect.
Consequently, the thermopower also approaches 
the two-orbital behavior with increasing $\Delta_\lambda$.

The local DOS for $\Delta_{\lambda}/\Gamma=0.05$ is shown 
in Fig. \ref{fig:DOS}(b).
At $\Delta_{B}/\Gamma=0.05$,
we can see that 
the value of the DOS around the Fermi level develops
and the Kondo resonance is located above the Fermi level,
which results in the enhancement of the conductance and
the negative value of the thermopower.

We summarize the main points of transport properties 
due to the three-orbital Kondo effect with 
two electrons as follows. 
In the absence of the $\lambda$-perturbation, 
the maximum structure of the conductance 
due to the singlet-triplet Kondo mechanism 
becomes obscure, because the Kondo temperature enhanced by 
the three-fold orbital degeneracy is much larger than the others. 
Introducing the $\lambda$-perturbation 
revives the magnetic-field-enhanced Kondo effect, 
such as the singlet-triplet Kondo effect, 
which is seen in the magnetic-field dependence of 
the conductance and the thermopower. 
In addition, we demonstrate that 
the conductance and the thermopower behavior 
approach that of the two-orbital system 
in large limit of the $\lambda$-perturbation. 
Experimentally, 
the measurement of the conductance and the thermopower 
may be useful to 
investigate dynamical properties of the DOS around the Kondo resonance.

We finally make a brief comment on the system without the Hund-coupling $J=0$. 
It corresponds to the negligibly small Hund-coupling 
$J \ll T_K^{\Delta_B=\Delta_\lambda=0}$. 
Then,
we do not need to take account of the competition between
the Hund-coupling and the level-splitting.
Therefore, the Kondo temperature decreases monotonously
with increasing the magnetic field which lifts the orbital degeneracy.
Thus, the conductance and the absolute value of the thermopower
in the Kondo regime
decreases monotonously with magnetic fields.
In this case, 
the $\lambda$-perturbation 
does not affect qualitative behavior of transport properties, 
because the local ground state is always the singlet state
for the finite magnetic field, 
irrespective of the $\lambda$-perturbation.

\subsection{With three electrons; $n_{tot}=3$}
In this subsection, 
we study the three-orbital Kondo effect with three electrons 
($n_{tot}=3$): 
The center of the energy levels in eq. (\ref{eq:energylevel}) is 
$\varepsilon_c = -5U/2$. 
It is expected that 
the Kondo effect occurs for any values of 
the level-splitting and the Hund-coupling, 
which significantly differs from the case with two electrons ($n_{tot}=2$). 
Here, we investigate the modulation of the Kondo effect 
caused by change of the symmetry in the system or 
the competition between the Hund-coupling and the level-splitting. 
We set parameters as $U/\Gamma=10$ and $T/\Gamma=0.025$ 
for the Kondo regime.

\subsubsection{Effect of $\lambda$-perturbation}
\label{subsec:J0}
First, we discuss the case in the absence of the Hund-coupling, $J=0$. 
\begin{figure}[tbp]
\begin{center}
\includegraphics[width=.9\linewidth]{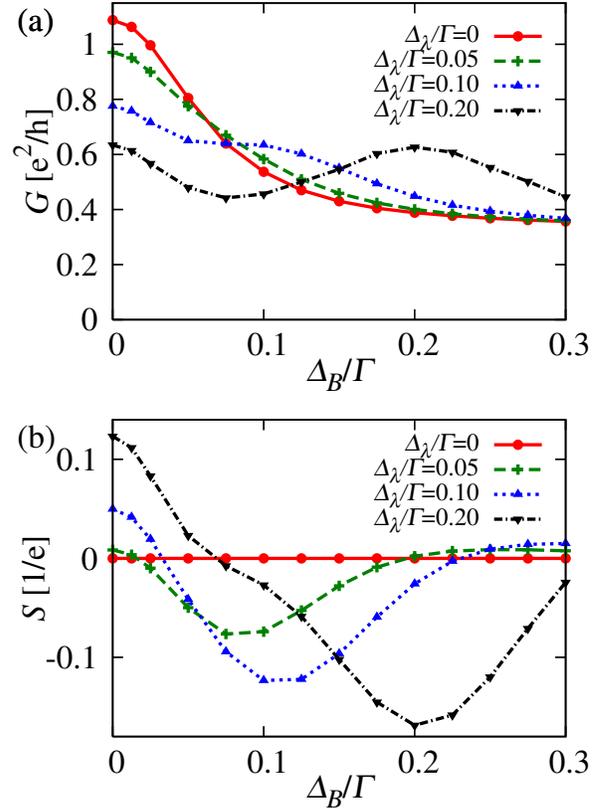}
\end{center}
\caption{(Color online) 
(a) The conductance $G$ and (b) the thermopower $S$
as a function of $\Delta_{B}$ for
$\Delta_{\lambda}/\Gamma=0, 0.05, 0.10, 0.20$
(solid, dashed, dotted, dashed-dotted line).
Parameters are set as $U/\Gamma=10$, $J=0$ and $T/\Gamma=0.025$. 
}
\label{fig:3e_realGS}
\end{figure}
\begin{figure}[tbp]
\begin{center}
\includegraphics[clip,width=.8\linewidth]{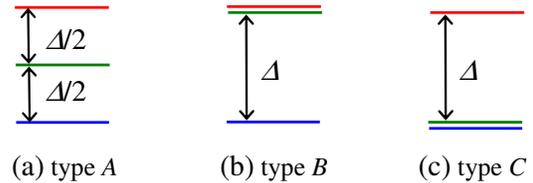} 
\caption{(Color online) 
Three selected configurations of the level-splitting scheme. 
$\Delta$ is a generic name of $\Delta_B$ and $\Delta_\lambda$. 
We label (a), (b) and (c) as type $A$, $B$ and $C$, respectively. 
In type $A$, each level-splitting is set $\Delta/2$ to compare with 
type $B$ and $C$. 
\label{fig:levelsplit}
}
\end{center}
\end{figure}

In Fig. \ref{fig:3e_realGS}, 
we show the conductance $G$ and the thermopower $S$ as a function of 
magnetic fields $\Delta_{B}$ for several choices of 
the $\lambda$-perturbation $\Delta_{\lambda}$. 
For small $\Delta_{\lambda}$ ($\Delta_{\lambda}/\Gamma=0$, 0.05), 
the conductance decreases monotonously 
with increasing the magnetic field. 
On the other hand, 
for large $\Delta_{\lambda}$ ($\Delta_{\lambda}/\Gamma=0.10$, 0.20), 
the conductance shows the nonmonotonous behavior 
as a function of $\Delta_{B}$. 
We explain the result by the 
change of the orbital-level configuration.

For $\Delta_{\lambda}=0$, 
triply-degenerated orbital levels 
are split only by the magnetic field. 
The conductance for the triply-degenerated orbital levels 
reaches $G=6e^2/h$ at absolute zero \cite{sakano06}, 
which results in the monotonous decrease of the conductance. 
On the other hand, by taking account of the $\lambda$-perturbation, 
the orbital-level configuration changes in a nontrivial way 
with increasing the magnetic field. 
In Fig. \ref{fig:levelsplit}, we show the 
some selected orbital configurations with the level-splitting $\Delta$, 
which denotes a generic name of the orbital splitting due to 
the magnetic field $\Delta_B$ and 
the $\lambda$-perturbation $\Delta_\lambda$. 
The orbital state at $\Delta_{B}=0$ 
corresponds to Fig. \ref{fig:levelsplit}(c), 
in which $\Delta=\Delta_\lambda$. 
When $\Delta_{B}$ increases to $\Delta_{B}=\Delta_{\lambda}/3$, 
the orbital state in Fig. \ref{fig:levelsplit}(a) is realized, 
in which $\Delta/2=\Delta_B=\Delta_{\lambda}/3$. 
With a further increase of $\Delta_{B}$, 
at $\Delta_{B}=\Delta_{\lambda}$, 
the orbital state in Fig. \ref{fig:levelsplit}(b) is realized, 
in which $\Delta=2\Delta_B=2\Delta_{\lambda}$. 

In the configuration described in Fig. \ref{fig:levelsplit}(a), 
while two electrons occupying the lowest orbital level 
do not contribute to the Kondo effect, 
the other electron causes the {\it SU}(2) Kondo effect 
with spin degrees of freedom. 
On the other hand, 
in the configuration described in Fig. \ref{fig:levelsplit}(b), 
the {\it SU}(4) Kondo effect 
with spin and orbital degrees of freedom becomes dominant. 
Note that the orbital configuration in Fig. \ref{fig:levelsplit}(c) is 
equivalent to Fig. \ref{fig:levelsplit}(b) 
via the particle-hole transformation. 
Thus, the dominant Kondo effect is modulated 
from the {\it SU}(4) Kondo effect through the {\it SU}(2) Kondo effect 
to the {\it SU}(4) Kondo effect
over the magnetic field. 
As mentioned above, 
the conductance of the {\it SU}(4) Kondo effect is larger 
than that of the {\it SU}(2) Kondo effect at finite temperatures, 
which results in the nonmonotonous magnetic-field dependence of 
the conductance. 

The characteristic magnetic-field dependence 
of the thermopower by the effect of the $\lambda$-perturbation 
is seen in Fig. \ref{fig:3e_realGS}(b). 
For the {\it SU}(4) Kondo effect, 
the effective filling is fewer than half-filling, 
which leads to the negative thermopower $S$, 
while the {\it SU}(2) Kondo effect results in $S \sim 0$. 
This results in the nonmonotonous behavior of the thermopower 
for the finite $\lambda$-perturbation ($\Delta_{\lambda}/\Gamma=0.05, 0.10, 0.20$). 
The fact the thermopower for $\Delta_{B} \sim \Delta_{\lambda}$ is 
opposite in sign to that for $\Delta_{B}=0$ 
corresponds to the particle-hole symmetric connection 
between the two regimes. 

We note that 
the conductance show the monotonous magnetic-field dependence 
for $\Delta_{\lambda}/\Gamma=0.05$, 
while we can see the nonmonotonous behavior in the thermopower 
even for $\Delta_{\lambda}/\Gamma=0.05$. 
The nonmonotonous magnetic-field dependence of the transport properties 
results from the modulation of the Kondo effect, as mentioned above.
It appears when 
$\Delta_\lambda$ smears the 20-fold Kondo effect 
due to the triply degenerate orbital state, 
$\Delta_\lambda > T_K^{\Delta_B=\Delta_\lambda=0}$, 
and the temperature is low enough to occur the {\it SU}(4) Kondo effect, 
$T < T_K^{SU(4)}$.
The order of the Kondo temperatures are estimated 
to be $T_K^{\Delta_B=\Delta_\lambda=0} \sim 0.1\Gamma$ and 
$T_K^{SU(4)} \sim 0.05\Gamma$ 
from the resonance width of the computed DOS. 
Then, the temperature of the system is not low enough 
than the Kondo temperature for the {\it SU}(2) Kondo effect 
estimated to be $T_K^{SU(2)} \sim 0.01\Gamma$, 
which also contributes to the nonmonotonous behavior of transport properties.
It is found that the thermopower more sensitively depends on 
the orbital configuration 
which changes the symmetry of the system and modulates the Kondo state 
than the conductance does. 

\subsubsection{Competition between Hund-coupling and level-splitting}
Let us now turn to investigate the effect of the Hund-coupling. 
As mentioned before, 
the Hund-coupling is competitive to the level-splitting 
as seen in the singlet-triplet Kondo effect. 
In particular, we consider the orbital configurations described in 
Figs. \ref{fig:levelsplit}(a) and \ref{fig:levelsplit}(b). 
We refer to them as type $A$ and type $B$ in the following, 
in which the {\it SU}(2) Kondo effect and the {\it SU}(4) Kondo effect 
are dominant, respectively.

\begin{figure}[tbp]
\begin{center}
\includegraphics[clip,width=.9\linewidth]{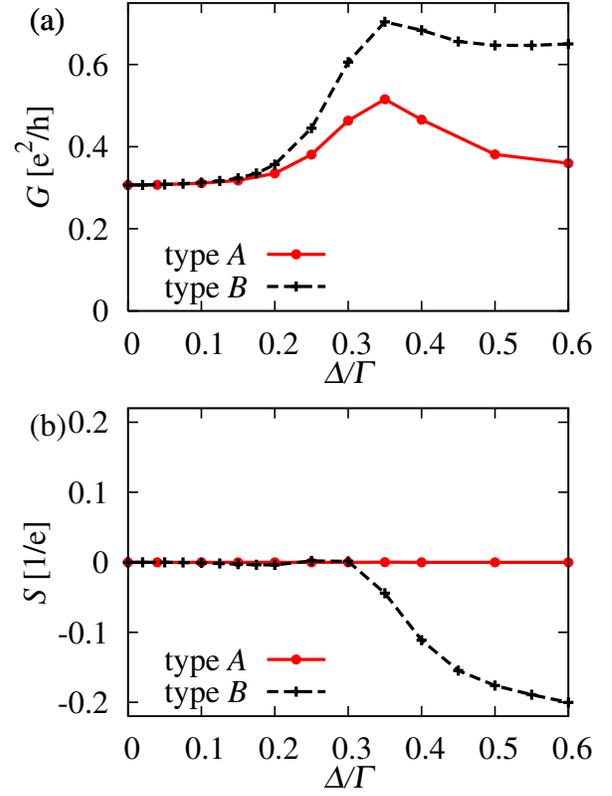}
\caption{
(Color online) 
(a) The conductance $G$ and (b) the thermopower $S$
as a function of $\Delta$ for the configurations type $A$ and $B$ 
with the Hund-coupling $J/\Gamma=0.5$. 
Other parameters are the same as in Fig. \ref{fig:3e_realGS}.
}
\label{fig:3eGS}
\end{center}
\end{figure}

In Fig. \ref{fig:3eGS}, 
we show the conductance $G$ and the thermopower $S$ 
as a function of the level-splitting $\Delta$ 
for the configurations type $A$ and $B$ 
with the Hund-coupling $J/\Gamma=0.5$. 
We explain the competition between the level-splitting $\Delta$ and 
the Hund-coupling $J$ from the ground state of the local Hamiltonian $H_d$. 
For small level-splittings $\Delta$, 
the quartet state, in which 
three electrons align their spins by the Hund-coupling $J$, 
is dominant for both type $A$ and $B$. 
As increasing $\Delta$, 
the {\it SU}(2) Kondo effect and {\it SU}(4) Kondo effect 
are dominant for type $A$ and $B$, respectively. 
For both types, the degeneracy increases at $\Delta = 3J/4(=0.375\Gamma)$, 
which results in the enhancement of the Kondo temperature 
at $\Delta \sim 3J/4$. 
This leads to the conductance maximum 
at $\Delta/\Gamma \sim 3J/4\Gamma(=0.375)$ for both types 
as seen in Fig. \ref{fig:3eGS}(a). 
For small level-splittings ($\Delta/\Gamma\le 0.15$), 
the Hund-coupling makes 
the distinction between orbital-splitting configurations unclear. 

One can deduce the magnetic-field dependence of transport properties 
from the above investigations on the effect of the $\lambda$-perturbation 
and Hund-coupling. 
For small magnetic fields and small $\lambda$-perturbations, 
the Hund-coupling makes the conductance and the thermopower 
hardly influenced by magnetic fields. 
Here, the thermopower is almost zero. 
On the other hand, when the magnetic field 
and/or the $\lambda$-perturbation becomes so large that 
the Hund-coupling does not work any more, 
the conductance and the thermopower show 
the nonmonotonous magnetic-field dependence 
by the effect of the $\lambda$-perturbation. 
Between these regimes the conductance may be enhanced 
by increasing the degeneracy. 
The negligibly small $\lambda$-perturbation 
$\Delta_\lambda \ll T_K^{\Delta_B=\Delta_\lambda=0}$ and 
Hund-coupling $J \ll T_K^{\Delta_B=\Delta_\lambda=0}$ does not affect 
transport properties. 

\subsection{Temperature dependence}
We have carried out our analysis by applying the NCA to even lower temperatures,
although this method is valid for temperatures 
around and higher than the Kondo temperature. 
In this subsection, 
we complementally investigate the temperature dependence of transport quantities 
in the typical parameter region, 
to confirm the reliability of our analysis. 

\begin{figure}[tbp]
\begin{center}
\includegraphics[width=.9\linewidth]{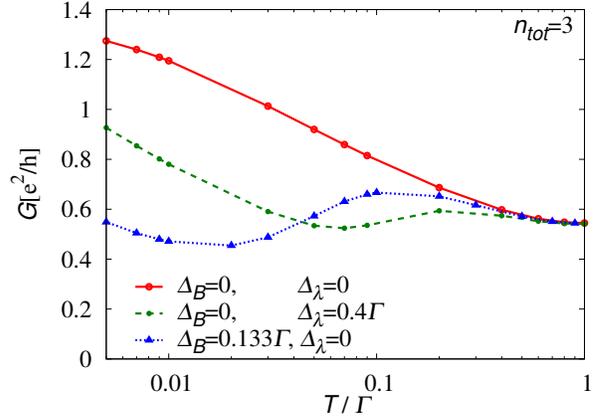}
\caption{(Color online) 
The conductance $G$ as a function of the temperature $T$ 
for $n_{tot}=3$ and 
$\Delta_B/\Gamma=\Delta_\lambda/\Gamma=0$ (solid line), 
$\Delta_B/\Gamma=0,\Delta_\lambda/\Gamma=0.4$ (dashed line), 
$\Delta_B/\Gamma=0.133,\Delta_\lambda/\Gamma=0$ (dotted line). 
Other parameters are set as $U/\Gamma=10$, $J/\Gamma=0$. 
}
\label{fig:Tdep}
\end{center}
\end{figure}

In Fig. \ref{fig:Tdep}, we show the temperature dependence of 
the conductance for three typical orbital configurations in $n_{tot}=3$ regime. 
We set the Hund-coupling $J=0$. 
For $\Delta_B = \Delta_{\lambda} = 0$, 
the conductance is enhanced by the 20-fold Kondo state around 
$T \sim T_K^{\Delta_B = \Delta_\lambda = 0} \sim 0.1 \Gamma$ and 
monotonously increases as the temperature decreases. 
The Kondo temperature for the 20-fold state is expected to be rather high 
in the parameter regime studied in this paper. 
Therefore, 
it is convenient to examine the low-temperature behavior described by 
the NCA in this parameter. 

Next we consider the case with the orbital level-splittings, 
$\Delta_B = 0, \Delta_{\lambda} = 0.4\Gamma$ 
($\Delta_B = 0.133\Gamma, \Delta_{\lambda} = 0$), 
where the {\it SU}(4) ({\it SU}(2)) Kondo state is realized 
in the large limit of the level-splittings. 
Around $T \sim 0.1 \Gamma$, 
the 20-fold Kondo state occurs due to large thermal fluctuation. 
However, at lower temperatures, the orbital splitting becomes effective, 
which results in the suppression of the conductance. 
At further lower temperatures $T \sim T_K^{SU(4)} \sim 0.05 \Gamma$ ($T_K^{SU(2)} \sim 0.01 \Gamma$), the {\it SU}(4) ({\it SU}(2)) Kondo state enhances the conductance again. 
Thus, we find that 
the qualitative behavior of the temperature dependence of the conductance is reasonable 
in the temperature range shown in Fig. \ref{fig:Tdep}. 

We note that 
the unitary limit of the conductance for the 20-fold Kondo state is $6e^2/h$, 
while the conductances for {\it SU}(4) and {\it SU}(2) Kondo states 
approach $2e^2/h$ \cite{sakano06}. 
Even at low temperatures in our calculation, 
the conductance for the 20-fold Kondo state does not seem 
to approach the unitary $6e^2/h$, 
but is always larger than that for the case with orbital-splittings. 
In addition, at low temperatures, 
the conductance for the {\it SU}(4) Kondo state is larger than 
that for the {\it SU}(2) Kondo state, which results from the fact that 
the Kondo temperature for the {\it SU}(4) Kondo state is larger than 
that for the {\it SU}(2) Kondo state, 
$T_K^{SU(4)} >T_K^{SU(2)}$. 

Therefore, we conclude that {\it quantitative} discussion by the NCA is no longer valid at lower temperatures than the Kondo temperature, but {\it qualitative} one is still enough reasonable in our temperature range. 
We can similarly expand this qualitative validity to other parameters, orbital configurations and the number of electrons.

\section{Summary}
\label{sec:sum}
We have studied the Kondo effect and transport properties
in vertical QD systems with orbital degrees of freedom.
By applying the NCA to the three-orbital Anderson impurity model
with the finite Coulomb interaction and Hund-coupling,
we have investigated 
the magnetic-field dependence of transport properties,
the conductance and the thermopower.
To consider the realistic situation in vertical QD systems,
we have introduced the $\lambda$-perturbation,
which is an additional orbital splitting caused by many-body effects
in the QD. 

With one electron contained in the three-fold orbitals of the QD,
the Hund-coupling hardly affects the Kondo effect. 
Then, the system crossovers from the three-orbital Kondo regime
to the two-orbital Kondo regime, 
as increasing the $\lambda$-perturbation. 
For small $\lambda$-perturbations, the conductance shows the plateau, 
which is characteristic behavior of the three- or more-orbital Kondo effect.
This has been experimentally confirmed in the vertical QD system with
well-tuned orbitals \cite{amaha}.

With the plural number of electrons,
the competition between the Hund-coupling and the orbital splitting
plays an important role.
When two electrons are contained 
in the triply degenerate orbitals, 
the maximum structure due to the singlet-triplet Kondo mechanism 
becomes obscure, because the Kondo temperature enhanced by 
the three-fold orbital degeneracy is very high. 
Introducing the $\lambda$-perturbation, the conductance maximum revives. 
For large $\lambda$-perturbations, 
the system is regarded as the two-orbital model and  
the conductance maximum appears due to singlet-triplet Kondo mechanism. 
For small $\lambda$-perturbations, 
we have clarified the significant difference 
in its mechanism from the two-orbital singlet-triplet Kondo effect. 
In this case, 
the conductance maximum appears 
not by the two-orbital singlet-triplet mechanism 
but by tangled three orbitals. 

With three electrons contained in the three-fold orbitals of the QD,
the change in the symmetry of the system 
with the magnetic field and the $\lambda$-perturbation
produces crucial effects on the transport properties.
Namely, the change of the symmetry causes 
the change of the dominant Kondo effect such as the $SU(2)$ and 
$SU(4)$ Kondo effect, 
which gives rise to the nonmonotonous magnetic-field 
dependence of the conductance and the thermopower.
Introducing the Hund-coupling, 
the singlet-triplet-type conductance maximum is induced by 
the competition between the Hund-coupling and the orbital splitting.  
For the strong Hund-coupling, transport properties 
hardly depend on the magnetic field and $\lambda$-perturbation. 

We have also discussed the temperature dependence of the Kondo effect 
and transport properties, although we have mainly studied the 
$\Delta_B$-dependence at a fixed finite temperature. 
At a fixed temperature $T$, 
the behavior of the transport properties $G$ and $S$ 
for varying parameters $\Delta_B$ and $\Delta_\lambda$ 
strongly depends on the relation between $T$ and the Kondo temperature $T_K$. 
Therefore, we have estimated $T_K$ from DOS and 
have compared it with $T$ for each $\Delta_B$ and $\Delta_\lambda$. 
In addition to $T_K$, 
absolute values of $G$ and $S$ change 
depending on the filling and the number of orbitals. 
In some specific situations, such as large limit of the level-splitting, 
$G$ and $S$ at absolute zero are exactly obtained 
by Bethe-ansatz solution \cite{sakano06,sakano07_proceedings}. 
Comparing our results with the exact values, we have discussed 
the qualitative behavior at lower temperatures.

The similar discussion can be applied to the 
multiorbital case which has more than three orbital degrees of freedom. 
The Kondo temperature, the unitary limit and 
the effective filling contribute significantly to the Kondo effect and 
related transport properties. 
The $\lambda$-perturbation causes the modulation of transport quantities in the unitary limit
and the Kondo temperature, which essentially affects 
the multiorbital Kondo effect and the transport properties. 
When the plural number of electrons are contained in the QD,  
the competition between the Hund-coupling and the orbital splitting 
induces a drastic change of the Kondo temperature. 
Although the thermopower is sensitive to the profile of the Kondo resonance
near the Fermi level, 
its behavior can be understood by considering the effective filling as above mentioned. 

As discussed in this paper, 
transport properties due to the multiorbital Kondo effect 
exhibit a variety of properties depending on the strength of 
the magnetic field, $\lambda$-perturbation, Hund-coupling, etc. 
The conductance and the thermopower reflect 
the strength and the location of the Kondo resonance, respectively. 
We expect that various properties due to the multiorbital Kondo effect 
are experimentally observed in the vertical QD 
by the measured conductance and thermopower, 
which gives detailed information on the electronic states in the QD. 

\section*{Acknowledgment}
The authors thank S. Tarucha, M. Eto, S. Amaha, Y. Tokura and N. Kawakami for useful comments and valuable discussions. 
Numerical computations have been done using the Yukawa Institute Computer Facility, Kyoto University, and the facilities of the Supercomputer Center, Institute for Solid State Physics, University of Tokyo.
TK and RS were supported by the Japan Society for the Promotion of Science. 
This work was partly supported by a Grant-in-Aid (No. 20540390) for 
Scientific Research from the Ministry of Education, Culture, Sports, 
Science, and Technology, Japan.

\end{document}